\documentclass[aps,prl,twocolumn,superscriptaddress,showpacs]{revtex4-1}
\usepackage[dvips]{graphicx}
\usepackage{multirow,wasysym}
\DeclareTextSymbol{\degre}{T1}{6}
\DeclareTextSymbol{\degre}{OT1}{23}

\begin{document}

\title{Pressure induced renormalization of energy scales in the unconventional superconductor FeTe$_{0.6}$Se$_{0.4}$}

\author{K. Marty}
\author{A. D. Christianson}
\author{A. M. dos Santos}
\author{B. Sipos}
\affiliation{Oak Ridge National Laboratory, Oak Ridge, Tennessee 37831 USA}

\author{K. Matsubayashi}
\author{Y. Uwatoko}
\affiliation{Institute for Solid State Physics, The University of Tokyo, Kashiwanoha, Kashiwa, Chiba 277-8581 Japan}
\affiliation{JST, TRiP, 5 Sanbancho, Chiyoda, Tokyo 102-0075 Japan}
\author{J. A. Fernandez-Baca}
\author{C. A. Tulk}
\author{T. A. Maier}
\author{B. C. Sales}
\author{M. D. Lumsden}
\affiliation{Oak Ridge National Laboratory, Oak Ridge, Tennessee 37831 USA}

\date{\today}

\begin{abstract}

We have carried out a pressure study of the unconventional superconductor FeTe$_{0.6}$Se$_{0.4}$ up to 1.5 GPa by neutron scattering, resistivity and magnetic susceptibility measurements. We have extracted the neutron spin resonance energy and the superconducting transition temperature as a function of applied pressure. Both increase with pressure up to a maximum at $\approx$1.3 GPa. This analogous qualitative behavior is evidence for a correlation between these two fundamental parameters of unconventional superconductivity. However, T$_c$ and the resonance energy do not scale linearly and thus a simple relationship between these energies does not exist even in a single sample. The renormalization of the resonance energy relative to the transition temperature is here attributed to an increased hybridization. The present results appear to be consistent with a pressure-induced weakening of the coupling strength associated with the fundamental pairing mechanism.

\end{abstract}



\maketitle


Unconventional superconductors like cuprates, heavy-fermion compounds, iron pnictides and chalcogenides all share some notable features. Perhaps the most salient is the presence of static or dynamic magnetism throughout the superconducting region of the phase diagram \cite{Hufner08, LumsdenChristianson10}.  A hallmark of this is a collective spin excitation that appears as a peak in the imaginary part of the dynamic susceptibility $\chi$''(\textbf{Q},$\omega$) \cite{Rossat-Mignod91, Fong99, Dai00, He02, Sato01, Stock08, Christianson08, Lumsden09, Qiu09, Park11}, often called the spin resonance. This resonance is localized in both wave vector and energy transfer, and its intensity is strongly coupled with the superconducting transition temperature. Although still open to interpretation, a commonly held view is that the spin resonance originates from a sign change of the superconducting order parameter on different parts of the Fermi surface \cite{Bulut96}. Within this picture, its existence is definitive evidence of unconventional superconductivity \cite{Eschrig06}.
The observation of the resonance signal in iron superconductors \cite{Christianson08, Lumsden09, Qiu09, Park11} provided further stimulus to explore the relationship between the resonance energy, $\omega_r$, and other characteristic energy scales such as the superconducting transition temperature, T$_c$, or the superconducting gap, $\Delta$ \cite{Hufner08, Paglione10, Yu09, Inosov11}.   

Recent studies show that there is ambiguity in interpreting the relationship between T$_c$ and $\omega_r$ \cite{Yu09,Inosov11}, in large part due to the difficulty in separating the intrinsic and extrinsic effects of chemical doping, such as disorder, inhomogeneity, and the influence of static magnetic order.  Consequently, a clean tuning parameter such as pressure has the potential to avoid these complications and yield further insight into the relationship of the resonance and unconventional superconductivity. Unfortunately, pressure dependent inelastic neutron scattering measurements are notoriously difficult and to date we are unaware of any reported studies of the spin resonance as a function of applied pressure.

The FeTe$_{1-x}$Se$_x$ family is a good candidate for such studies, as large single crystals can be grown and T$_c$ shows a substantial sensitivity to applied pressure \cite{Mizuguchi08, Horigane09}. In particular, for compositions close to $x\approx0.5$, the samples do not exhibit long-range magnetic order and T$_c$ increases with pressure reaching a maximum at $\approx$2GPa, a pressure amenable to a number of experimental techniques. 
In this letter, we present inelastic and elastic neutron scattering, resistivity and magnetic susceptibility measurements of FeTe$_{0.6}$Se$_{0.4}$ (T$_C\approx$12K) up to 1.5 GPa. T$_c$ and $\omega_r$ show a similar qualitative behavior, although $\omega_r$ does not increase as much as T$_c$ as a function of pressure, revealing the lack of propotionality between these two energies. The renormalization of $\omega_r$ relative to T$_c$ is explained by a pressure-induced increase of orbital overlap reducing the strength of the correlations in the sample. Within the model of sign change of the superconducting order parameter, this result implies a decrease of the superconducting pairing strength as a function of pressure.

The FeTe$_{0.6}$Se$_{0.4}$ crystal studied here was grown using a modified Bridgman technique \cite{Sales09}. The stoichiometry was determined by energy dispersive x-ray analysis, resulting in 1.02$\pm$0.02 for Fe, 0.6$\pm$0.02 for Te and 0.4$\pm$0.02 for Se. The inelastic neutron scattering experiments were performed on the HB-3 triple axis spectrometer at the High Flux Isotope Reactor of ORNL with collimations of 48'-60'-80'-120'. A McWhan piston-cylinder pressure cell \cite{Onodera87} was used with 3M FC-75 fluorinert as the pressure medium. A 0.5 g crystal was encapsulated in the inner BeCu neutron pressure cell ($\diameter$*h=5mm*10mm) with the [1-10] direction vertical.
Room temperature neutron powder diffraction was performed using the SNAP diffractometer at the Spallation Neutron Source at ORNL. NaCl powder was ground into a 0.8 g FeTe$_{0.6}$Se$_{0.4}$ sample (from the original single crystal) to use as a pressure calibration standard. The sample was loaded into a Paris-Edinburgh press fitted with single toroid cubic boron nitride anvils with the incident beam through the TiZr null scattering alloy gasket. Pressure was determined by application of the isothermal NaCl equation of state determined by Decker \cite{Decker71} to the refined lattice parameters obtained at each measured pressure from LeBail fits using the GSAS software suite \cite{Larson00}.  
High pressure resistivity measurements were performed in an easyLab Mcell 30. The electrical contacts were made using Dupont 4929N silver paste. The pressure was determined during the pressurization at room temperature with calibrated manganin wire and was also calculated from the applied load. Fluorinert (FC-75) was used as the pressure medium. 
The DC magnetization was measured by a commercial SQUID magnetometer (MPMS) in BeCu piston cylinder cell using Daphne7373 as the pressure medium.

\begin{figure}
		\includegraphics[width=1.0\columnwidth]{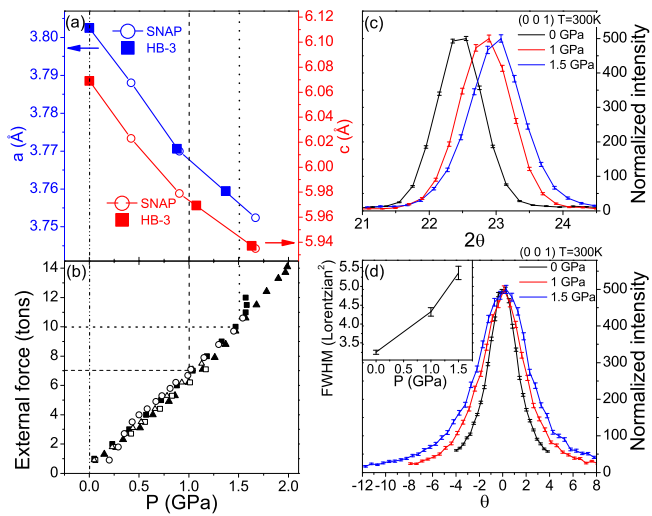}
		\caption{(Color online) (a) Pressure dependence of the FeTe$_{0.6}$Se$_{0.4}$ lattice parameters $a$ (in blue) and $c$ (in red) measured by neutron powder diffraction on the time-of-flight diffractometer SNAP (open circles), and single crystal neutron diffraction on the triple axis spectrometer HB-3 (squares). (b) Pressure calibration of the McWhan pressure cell as a function of the applied external force. (c) Pressure dependence of the $\theta-2\theta$ scan on (0 0 1). (d) Evolution of the single crystal mosaic as a function of pressure. Inset: Lorentzian squared FWHM as a function of pressure.}
		\label{calibration}
\end{figure}

According to the different calibration tests for the McWhan pressure cell \cite{Onodera87} (by measuring the resistance of a manganin wire under pressure), the chosen external forces applied on the pressure cell should result in room temperature pressures of $\approx$0, 1 and 1.5 GPa (fig.\ref{calibration}(b)). The lattice parameters of FeTe$_{0.6}$Se$_{0.4}$ were extracted from $\theta-2\theta$ scans through the (1 1 0) and (0 0 1) Bragg reflections (see fig.\ref{calibration}(a) and (c)) at room temperature on the HB-3 spectrometer for the three different chosen pressure points (applied forces of respectively 1, 7 and 10 tons), and their relative decrease compared to the absolute values of these same lattice parameters obtained from neutron powder diffraction on the SNAP diffractometer. For this purpose, the first pressure point, ambient-pressure/1 ton, was considered as a common reference and the HB-3 lattice parameters were normalized to the SNAP ones. The measured relative changes (from HB-3) were then were then compared to the lattice parameters extracted from the SNAP data (see fig.\ref{calibration}(a)). While the lattice parameter $a$ shows higher values than expected for these pressures, the lattice parameter $c$ shows slightly lower values, which suggests that the actual pressures at room temperature are close to the expected ones.
One of the first visible effects of applied pressure, besides the reduction of the lattice parameters, is the broadening of the crystal mosaic. The rocking curves show a significant increase in their width (see fig.\ref{calibration}(d) and inset). A Lorentzian-squared function provides a better description of the rocking curves than a classical Gaussian function. 

\begin{figure}
		\includegraphics[width=1.0\columnwidth]{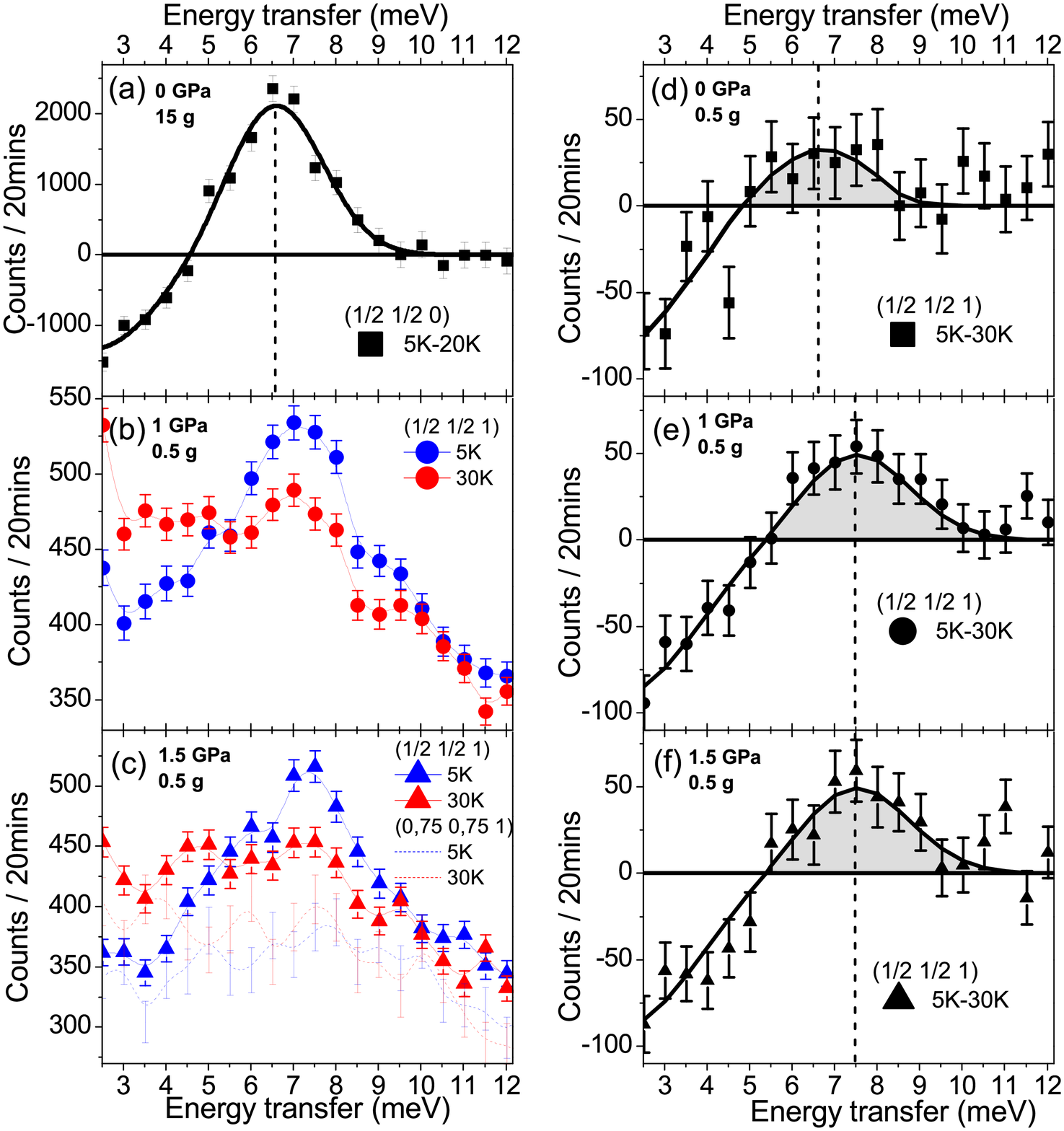}
		\caption{ (Color online) Solid lines are guide for the eye. (a) Spin resonance in a 15 g single crystal of FeTe$_{0.6}$Se$_{0.4}$ with no applied pressure. Each point was measured for $\approx$7 mins. The resonance energy is indicated by a vertical dotted line. (b)-(f) All data were measured on a 0.5 g single crystal of FeTe$_{0.6}$Se$_{0.4}$  in the pressure cell and normalized to 20 mins/point. (b) and (c) Constant-Q scans at (1/2 1/2 1) measured below T$_c$ (5K, blue) and above T$_c$ (30K, red) at respectively 1 GPa (circles, 80mins/point) and 1.5 GPa (triangles, 60 mins/point). At 1.5 GPa, background constant-Q scans (dotted lines) measured at (0.75 0.75 1) for 20 mins/point are shown for T=5 K (blue) and T=30 K (red). (d) Spin resonance at 0 GPa. Each point was measured for 40 mins. The resonance energy, the same as (a), is indicated by a vertical dotted line. (e) Same as (d) for 1 GPa and 80 mins/point. (f) Same as (d) and (e) for 1.5 GPa and 60 mins/point. The line and shaded area are the same as 1 GPa to emphasize the similar spectral weight and resonance energy for both pressures.}
		\label{neutrons}
\end{figure}

In this FeTe$_{1-x}$Se$_x$ family, the neutron spin resonance has been shown to be two dimensionnal, centred at a \textbf{Q} of (1/2 1/2 L) where L indicates the irrelevant direction \cite{Qiu09}. Above T$_c$, the spin excitations in FeTe$_{1-x}$Se$_x$ originate from an incommensurate wave vector near (1/2 1/2 L) \cite{Lumsden10}.  Below T$_c$ there is a suppression of low energy spectral weight transferred to higher energy resulting in the appearance of a resonance peak at $\omega_r\approx$6.5 meV. 
Constant-Q scans at (1/2 1/2 L) were measured on the FeTe$_{0.6}$Se$_{0.4}$ crystal at 0, 1, and 1.5 GPa. The same, superconductivity induced redistribution of spectral weight can be seen in the constant-Q scans under applied pressure, as seen on fig.\ref{neutrons}(b) for 1 GPa and fig.\ref{neutrons}(c) for 1.5 GPa.   In each case there is a clear additional signal in the inelastic spectrum corresponding to the spin resonance.  While the background introduced by the pressure cell is substantial, it is not insurmountable and is indicated in fig.\ref{neutrons}(c).   

The resonance signal is obtained by subtracting the data collected above T$_c$ from the data collected below T$_c$. This procedure results in a positive difference centered at the resonance energy $\omega_r$ and an associated negative difference at lower energy corresponding to the opening of a gap in the spin excitation spectrum. Figure \ref{neutrons}(a) displays this signal measured on a large single crystal of the same concentration outside of the pressure cell indicating the expected resonance at $\omega_r$=6.6 meV.  For comparison, fig.\ref{neutrons}(d) shows the results of similar measurements on the 0.5 g sample within the pressure apparatus.  The effect of applied pressure on the resonance is illustrated by figures \ref{neutrons}(e) and (f) (1 and 1.5 GPa respectively). First, $\omega_r$ is shifted to a higher energy at 1 GPa ($\omega_r$=7.5 meV), but does not increase further at higher pressure. Second, the spectral weight enhancement below T$_c$ follows the same qualitative behavior as $\omega_r$ with a clear increase at 1 GPa and does not increase further at 1.5 GPa.

\begin{figure}[t]
	\begin{center}
		\includegraphics[width=1.0\columnwidth]{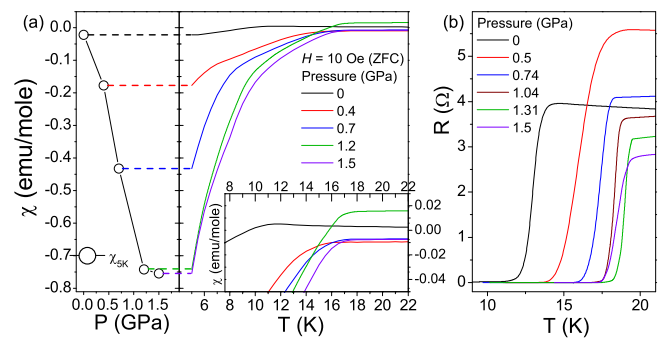}
		\caption{ (Color online) (a) Temperature dependence of magnetic susceptibility $\chi_{mag}$ for various applied pressure in the 0-1.5 GPa range. Inset: zoom on $\chi_{mag}$(T) in the superconducting transition zone. Left panel: Lowest value of $\chi_{mag}$, proportional to the screening effect in the sample. (b) Temperature dependence of electrical resistivity for various applied pressure in the 0-1.5 GPa range. }
		\label{TcvsP}
	\end{center}
\end{figure}

\begin{figure}
		\includegraphics[width=0.8\columnwidth]{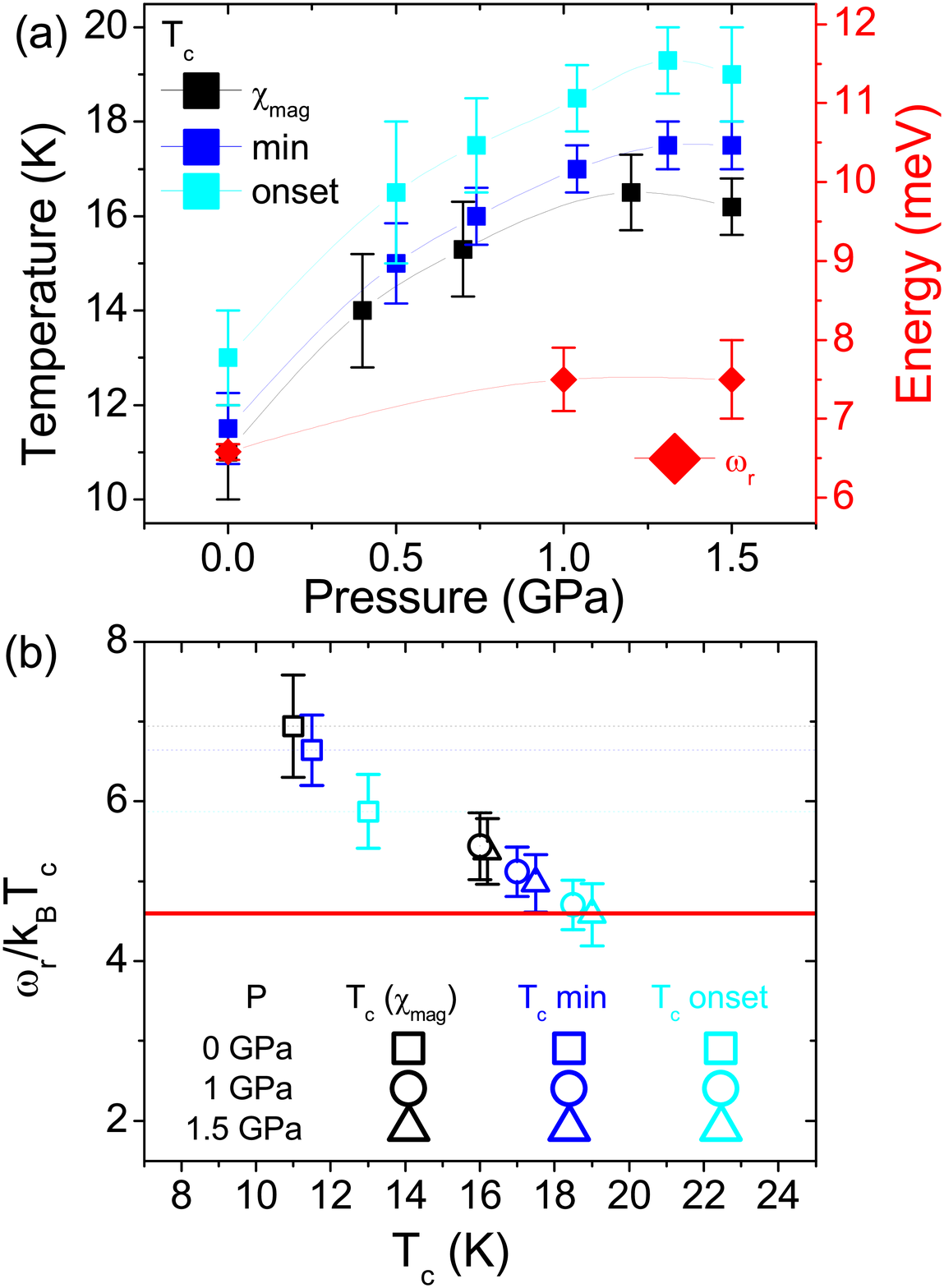}
		\caption{ (Color online) (a) Pressure dependence of $\omega_r$ (in red) and T$_c$ as defined by the magnetic susceptibility (in black), and the resistivity (T$_c$ min  in dark blue, T$_c$ onset in light blue). (b) T$_c$ dependence of the $\omega_r$/k$_B$T$_c$ ratio of FeTe$_{0.6}$Se$_{0.4}$ for all pressures (squares for 0 GPa, circles for 1 GPa and triangles for 1.5 GPa), and all definitions of the transition temperature. The horizontal dotted lines correspond to the value of $\omega_r$/k$_B$T$_c$ at 0 GPa for all three definitions of T$_c$. The horizontal red line corresponds to $\omega_r$/k$_B$T$_c$=4.6 as previously observed for iron superconductors \cite{Paglione10, Yu09}. }
		\label{ErvsTc}
\end{figure}

The large thermal mass of the pressure cell combined with long counting times prohibited an accurate determination of T$_c$ from the onset of the resonance.  Therefore the pressure dependence of T$_c$ was determined from resistivity and magnetic susceptibility measurements on samples from the same growth as the sample used for the inelastic neutron scattering measurements (see fig.\ref{TcvsP}). Although different methods are commonly used to extract T$_c$ from resistivity data, for completeness we adopt two metrics: (1) the onset of the drop in resistivity due to the appearance of superconductivity (T$_c$ onset) and (2) the temperature at which the resistivity achieves the minimum value (T$_c$ min).  For magnetic susceptibility data, we define T$_c$ as the temperature where $\chi_{mag}$ starts to decrease. Here, as shown in fig.\ref{ErvsTc}(a), T$_c$ obtained from all three methods show similar qualitative behavior despite a constant offset in absolute value.  T$_c$ rises with pressure to a maximum at $\approx$1.3 GPa, then decreases at higher pressures. Similar behavior was observed in other members of the FeTe$_{1-x}$Se$_x$ family \cite{Mizuguchi08, Horigane09}. The lowest value of $\chi_{mag}$, at the lowest temperature (5 K), is a manifestation of the screening effect in the sample. $\chi_{5K}$ drops with applied pressure reaching a plateau above $\approx$1 GPa (see fig.\ref{TcvsP}(a) left). This increased screening effect could indicate an enhancement of the superconducting volume fraction with pressure which is consistent with the increase in intensity observed in the inelastic neutron scattering measurements (figs.\ref{neutrons}(d)-(f)). It is tempting to correlate the increase in resonance spectral weight and increasing T$_c$ but the potential change in volume fraction makes such comparisons impossible.

Figure \ref{ErvsTc}(a) displays a comparison of T$_c$(P) and $\omega_r$(P), where $\omega_r$ is expressed in units of Kelvin and scaled by a constant to coincide with the lowest estimate of T$_c$(P) at ambient pressure. Despite a similarity in the behavior of $\omega_r$(P) and T$_c$(P), the resonance energy does not increase nearly as rapidly as T$_c$. To further clarify this point, fig.\ref{ErvsTc}(b) shows the ratio $\omega_r$/k$_B$T$_c$ as a function of T$_c$.  Both figs.\ref{ErvsTc}(a) and (b) clearly demonstrate that $\omega_r$ is not proportional to T$_c$. This fact is independent of the definition we choose for T$_c$.  Here the power of this statement comes from the realization of this intrinsic observation in a single sample, pressure being the means to tune the superconducting properties without modifying the system, the composition, or without additional doping-induced chemical inhomogeneity. Applying pressure likely increases orbital overlap and hybridization, implying a more itinerant and less correlated system. Within this context, the observed renormalization of $\omega_r$ relative to T$_c$ reflects the impact of the electronic correlations on the energy scales relevant in unconventional superconductivity.

It is interesting to compare this observation to the data obtained from multiple 122 materials \cite{Inosov11}. These materials have a q$_z$ dependent resonance mode where the ratio, $\omega_r$/k$_B$T$_c$, is constant for q$_z$=$\pi$  and seems to vary with T$_c$ for q$_z$=0. Interestingly, the data presented in fig.\ref{ErvsTc}(b) is qualitatively similar to the q$_z$=0 data for the 122 compounds, suggesting that when 3-dimensional, the resonance relevant for superconductivity, and which should be compared to either T$_c$ or $\Delta$, might be the one measured at q$_z$=0.

This result also has more fundamental implications. Since the neutron spin resonance is often compared to the superconducting gap $\Delta$ \cite{Hufner08, Yu09, Inosov11}, it is interesting to examine the consequences of our results for the underlying pairing mechanism. Unfortunately, no data of $\Delta$ as a function of pressure is available. Therefore we discuss our results in the context of a simple Hubbard model and consider the aforementioned model of sign change of the superconducting order parameter\cite{Bulut96}. Within this context, a screened, on-site, intra-orbital Coulomb interaction, U, renormalizes $\omega_r$ to a value lower than the energy of the particle-hole continuum, 2$\Delta$ \cite{Mazin95, Manske01}. Pressure increases orbital overlap, effectively reducing U, and in turn increasing the ratio of $\omega_r$/2$\Delta$. In the very weak-coupling limit, we expect $\omega_r$/2$\Delta \approx$1 \cite{Mazin95, Manske01}. For either increasing or constant $\omega_r$/2$\Delta$, the observed decrease of $\omega_r$/k$_B$T$_c$ between 0 and 1 GPa implies a reduction of 2$\Delta$/k$_B$T$_c$. This is a fingerprint of a weakening of the superconducting pairing strength (even though paradoxically the energy scale T$_c$ has risen). 

To conclude, we have performed a pressure dependence study of the neutron spin resonance and superconducting transition temperature of an unconventional FeTe$_{0.6}$Se$_{0.4}$ superconductor up to 1.5 GPa. Free from any constraint induced by chemical substitution, we have shown that the resonance energy and T$_c$ are indeed correlated, although not by linear proportionality. This result is attributed to an increased hybridization induced by applied pressure, which in turn implies a weakening of the superconducting coupling strength.


\begin{acknowledgments}
Research at ORNL is sponsored by the Scientific User Facilities Division and the Materials Sciences and Engineering Division, Office of Basic Energy Sciences, U.S. DOE. Work in Japan was supported by a Grant-in-Aid for Research (Nos. 21340092, 20102007, and 19GS0205) from the Ministry of Education, Culture, Sports, Science, and Technology, Japan. We acknowledge discussions with D. J. Singh and D. Mandrus. We thank E. D. Specht for the use of the X-ray La$\ddot{u}$e and S. Kulan and J. J. Molaison for technical support.
\end{acknowledgments}


\end{document}